\documentclass{article}
\usepackage{amsmath,amsfonts,amssymb}
\usepackage{cite}
\usepackage{color}
\usepackage{algorithm}
\usepackage[noend]{algpseudocode}
\usepackage{epsfig}
\usepackage{hyperref}
\usepackage{authblk}

\begin{document}

\title{Accelerating eigenvector and pseudospectra computation using
  blocked multi-shift triangular solves}
\author[1]{Tim Moon}
\author[1,2]{Jack Poulson}
\affil[1]{Institute for Computational and Mathematical Engineering, Stanford University, CA, USA}
\affil[2]{Department of Mathematics, Stanford University, CA, USA}
\maketitle

\begin{abstract}
  Multi-shift triangular solves are basic linear algebra calculations
  with applications in eigenvector and pseudospectra computation. We
  propose blocked algorithms that efficiently exploit Level 3 BLAS to
  perform multi-shift triangular solves and safe multi-shift
  triangular solves. Numerical experiments indicate that computing
  triangular eigenvectors with a safe multi-shift triangular solve
  achieves speedups by a factor of 60 relative to LAPACK. This
  algorithm accelerates the calculation of general eigenvectors
  threefold. When using multi-shift triangular solves to compute
  pseudospectra, we report ninefold speedups relative to EigTool.
\end{abstract}

\section{Introduction}
A common task in numerical linear algebra is the design of algorithms
that can be efficiently implemented using the Basic Linear Algebra
Subroutines (BLAS). BLAS is a specification of low-level Fortran
routines for vector operations (Level 1) \cite{lawson1979basic},
matrix-vector operations (Level 2) \cite{dongarra1988extended,
  dongarra1988algorithm}, and matrix-matrix operations (Level 3)
\cite{dongarra1990set, dongarra1990algorithm}. It has been in
development since the late 1970s and there currently exist a multitude
of high-quality (and often vendor-tuned) implementations such as
OpenBLAS \cite{zhang2012model, wang2013augem}, ATLAS
\cite{whaley2005minimizing}, BLIS \cite{van2015blis}, and Intel
MKL. Leveraging BLAS, especially Level 3 BLAS, can achieve substantial
improvements in performance since it uses highly optimized code that
efficiently exploits a machine's memory architecture and processor
capabilities.

\paragraph{}
Despite a large body of work devoted to utilizing Level 3 BLAS in
high-quality linear algebra packages like LAPACK
\cite{anderson1999lapack}, there still exist important routines that
are limited to Level 2 BLAS. For instance, the calculation of
triangular eigenvectors in the LAPACK routine \texttt{xTREVC} is
dominated by safe triangular solves that employ Level 2 BLAS routines
\cite{gates2014accelerating}.  In addition, matrix pseudospectra are
typically computed with embarrassingly parallel Level 2 BLAS
triangular solves \cite{trefethen2005spectra}.  This paper proposes
blocked implementations for multi-shift triangular solves that address
both of these problems.

\section{Multi-shift Triangular Solve} Given an \(m\times m\) upper
triangular matrix \(U\) and right-hand side vectors
\(b_1,\cdots,b_n\), the triangular solve problem seeks solution
vectors \(x_1,\cdots,x_n\) that solve \(n\) triangular systems of the
form\footnote{This paper will focus on the left upper triangular
  case. However, similar techniques can be applied for the lower
  triangular and right matrix cases. For instance, the left lower
  triangular case requires forward substitution instead of back
  substitution.}
\begin{align}
  U x_j &= b_j.
\end{align}
Consolidating the \(b_j\)'s and \(x_j\)'s into \(m\times n\) matrices
\(B\) and \(X\), respectively, this is equivalent to solving the
matrix problem \(UX=B\).  Assuming \(U\) is well-conditioned, the
naive approach is to apply back substitution to each right-hand side:
\begin{algorithm}[H]
  \caption{Single triangular solve with back substitution}
  \label{algorithm:trsv}
  \begin{algorithmic}
    
    \Procedure{Trsv}{$U,b$}
    \Comment \(b\) is overwritten with \(x\)
    
    \For{$i=m:-1:1$}

    \State \(\mathcal{I}_0 := 1:i-1\)

    \State \(b(i) := b(i) / U(i,i)\) \Comment Diagonal step

    \State \(b(\mathcal{I}_0) := b(\mathcal{I}_0) - b(i) * U(\mathcal{I}_0,i)\) \Comment Substitution step (\texttt{xAXPY})

    \EndFor

    \EndProcedure

  \end{algorithmic}
\end{algorithm}
\noindent
Applying this algorithm to one right-hand side takes \(m\) divisions
in the diagonal step and \(\sim m^2 \) flops in the substitution step,
so the computational work is dominated by the substitution step if the
matrix dimension is sufficiently large. This routine is backward
stable and implemented in the Level 2 BLAS routine
\texttt{xTRSV}. However, given a block size \(n_b\), we can apply a
blocked algorithm \cite{golub2012matrix}:
\begin{algorithm}[H]
  \caption{Triangular solve with blocked back substitution}
  \label{algorithm:trsm}
  \begin{algorithmic}
    \Procedure{Trsm}{$U,B$}
    \Comment \(B\) is overwritten with \(X\)

    \For{$i = m-n_b+1:-n_b:1$} \Comment Assume \(m\) is a multiple of
    \(n_b\)

    \State \( \mathcal{I}_0 := 1:i-1 \)

    \State \( \mathcal{I}_1 := i:i+n_b-1\)
    
    \For{$j=1:n$}

    \State \Call{Trsv}{$U(\mathcal{I}_1,\mathcal{I}_1), B(\mathcal{I}_1,j)$} 
    \Comment Diagonal block step (\texttt{xTRSV})

    \EndFor

    \State \( B(\mathcal{I}_0,:) := B(\mathcal{I}_0,:) - U(\mathcal{I}_0,\mathcal{I}_1) * B(\mathcal{I}_1,:) \)
    \Comment Substitution step (\texttt{xGEMM})

    \EndFor

    \EndProcedure
  \end{algorithmic}
\end{algorithm}
\noindent
The blocked algorithm takes the same number of flops as the naive
algorithm, but it uses the Level 3 BLAS routine \texttt{xGEMM} to
perform matrix multiplication in the substitution step.  The fraction
of flops taking place in the substitution step (the ``level-3
fraction'') is approximately \(1-n_b/m\), so the calculation is
efficient if \(m\gg n_b\). This routine is implemented in the Level 3
BLAS routine \texttt{xTRSM}.

\paragraph{}
The multi-shift triangular solve problem is a variant of the standard
triangular solve problem.\footnote{We remark that the multi-shift
  Hessenberg solve problem has a very similar form, replacing the
  upper triangular matrix with an upper Hessenberg one. This problem
  can be solved by computing RQ factorizations and performing back
  substitution \cite{henry1994shifted, bischof1996parallel}. Blocked
  implementations are asymptotically dominated by Level 2 BLAS
  routines.} Given scalar shifts \(\lambda_1,\cdots,\lambda_n\),
we seek to solve \(n\) triangular systems of the form
\begin{align}
  \left(U-\lambda_j I\right) x_j &= b_j.
\end{align}
Each triangular system has a different matrix, so a naive approach is
to apply unblocked back substitution on each system. However, the
matrices only differ in the diagonal entries. This implies we can use
blocked back substitution with a modification to the diagonal block
step:
\begin{algorithm}[H]
  \caption{Multi-shift triangular solve with blocked back substitution}
  \label{algorithm:multi-shift trsm}
  \begin{algorithmic}
    \Procedure{MultiShiftTrsm}{$U, \lambda, B $}
    \Comment \(B\) is overwritten with \(X\)

    \For{$i = m-n_b+1:-n_b:1$} \Comment Assume \(m\) is a multiple of
    \(n_b\)

    \State \( \mathcal{I}_0 := 1:i-1 \)

    \State \( \mathcal{I}_1 := i:i+n_b-1\)
    
    \For{$j=1:n$}

    \State \Call{Trsv}{$U(\mathcal{I}_1,\mathcal{I}_1)-\lambda(j) * I, B(\mathcal{I}_1,j)$} 
    \Comment Diagonal block step (\texttt{xTRSV})

    \EndFor

    \State \( B(\mathcal{I}_0,:) := B(\mathcal{I}_0,:) - U(\mathcal{I}_0,\mathcal{I}_1) * B(\mathcal{I}_1,:) \)
    \Comment Substitution step (\texttt{xGEMM})

    \EndFor

    \EndProcedure
  \end{algorithmic}
\end{algorithm}
\noindent
As before, the bulk of the computation is performed efficiently with
Level 3 BLAS in the substitution step.

\section{Safe Multi-Shift Triangular Solve}
If the \(m\times m\) upper triangular matrix \(U\) is ill-conditioned
or singular, performing a triangular solve with back substitution may
result in division by zero or floating point overflow. To avoid these
pitfalls, we consider the (single) safe triangular solve
problem. Given a nonzero right-hand side vector \(b\), we seek a
nonzero solution vector \(x\) and a scaling factor \(s\) in the unit
interval \(\left[0,1\right]\) such that
\begin{align}
  U x &= s b.
\end{align}
The LAPACK routine \texttt{xLATRS} solves this problem with
safeguarded back substitution \cite{anderson1991robust}. This routine
begins by estimating the growth of entries during standard back
substitution. If we run \(j\) iterations of back substitution,
\(b(i)\) will be overwritten with \(x(i)\), where \(i=m-j+1\). At this
stage of the calculation we define \(M(i)=\left|b(i)\right|\) and
\(G(i)=\left\lVert b(1:i-1) \right\rVert_\infty\). We have the initial
values \(M(m+1)=0\) and \(G(m+1)=\left\lVert b \right\rVert_\infty\)
and the bounds
\begin{align}
  M(i)
  &\leq \frac{G(i+1)}{\left| U(i,i) \right|} \nonumber \\
  &\leq \max\left\lbrace \frac{G(i+1)}{\left|U(i,i)\right|}, M(i+1) \right\rbrace \\
  G(i)
  &\leq G(i+1) + M(i) \,\left\lVert U(1:i-1,i) \right\rVert_\infty \nonumber \\
  &\leq G(i+1) \,\left( 1 + \frac{\left\lVert U(1:i-1,i) \right\rVert_\infty}{\left| U(i,i) \right|} \right).
\end{align}
These bounds can be computed recursively\footnote{In practice, we
  compute lower bounds to the reciprocals \(1/M(i)\) and \(1/G(i)\) to
  avoid overflow.} and they increase monotonically with each
iteration. Thus, the worst-case growth can be estimated by computing
bounds for \(M(1)\) and \(G(1)\). If the growth is not too large,
i.e.\ less than a machine-dependent overflow constant \(\Omega\), then
we can confidently apply standard back substitution and set
\(s=1\). Otherwise, we must check at each step of back substitution to
avoid overflow and division by zero. In the \(j\)th iteration of back
substitution, we compute the following quantity prior to the diagonal
step,
\begin{align}
  M(i) &= \left| \frac{b(i)}{U(i,i)} \right|.
\end{align}
In the case \(U(i,i)=0\), we approximate \(U(i,i)\) with some nonzero
\(\delta=O(\epsilon \lVert U\rVert_\infty)\), where \(\epsilon\) is
machine epsilon. Note that this approximation does not disrupt the
backward stability of back substitution.  If \(M(i)\geq \Omega\), then
\(s\) is reduced until \(M(i) < \Omega\) to protect against numerical
issues in the diagonal step. After the diagonal step and before the
substitution step, we compute the bound
\begin{align}
  G(i)
  &\leq G(i+1) + M(i) \left\lVert U(1:i-1,i) \right\rVert_\infty.
\end{align}
If \(G(i)\geq \Omega\), then \(s\) is reduced until \(G(i) < \Omega\)
to protect against numerical issues in the substitution step.  The
complete algorithm is outlined below:
\begin{algorithm}[H]
  \caption{Single safe triangular solve with safeguarded back substitution}
  \label{algorithm:safe trsv}
  \begin{algorithmic}
    
    \Procedure{SafeTrsv}{$U,b,s$}
    \Comment \(b\) is overwritten with \(x\)
    
    \State \(s := 1\)
    
    \State \(M := 0\)
    \State \(G := \left\lVert b \right\rVert_\infty \)

    \For{$i=m:-1:1$}

    \State \( M := \max\left\lbrace G /\left|U(i,i)\right|, M \right\rbrace\)

    \State \( G := G * \left( 1 + \left\lVert U(1:i-1,i) \right\rVert_\infty/\left| U(i,i) \right| \right) \)

    \EndFor
    
    \If{$M<\Omega$ and \(G<\Omega\)}

    \State \Call{Trsv}{$U,b$} \Comment \texttt{xTRSV}

    \Else

    \State \(G := \left\lVert b \right\rVert_\infty\)

    \For{$i=m:-1:1$}

    \State \(\mathcal{I}_0 := 1:i-1\)

    \State \(M := \left|{b(i)}/{U(i,i)}\right|\) \Comment{Assume
      \(U(i,i)\neq 0\)}

\If{$M \geq
      \Omega$} 

    \State Choose \(t\in\left(0,1\right)\) so that \(t* M< \Omega\)

    \State \(b := t * b\) \Comment \texttt{xSCAL}

    \State \(s := t * s\)

    \State \(M := t * M\)

    \State \(G := t * G\)

    \EndIf

    \State \(b(i) := b(i) / U(i,i)\) \Comment Diagonal step

    \State \(G := G + M * \left\lVert U(1:i-1,i) \right\rVert_\infty\)

    \If{$G \geq \Omega $}

    \State Choose \(t\in\left(0,1\right)\) so that \(t *G < \Omega\)

    \State \(b := t *b\) \Comment \texttt{xSCAL}

    \State \(s := t *s\)

    \State \(M := t*M\)

    \State \(G := t *G\)
    
    \EndIf

    \State \(b(\mathcal{I}_0) := b(\mathcal{I}_0) - b(i) * U(\mathcal{I}_0,i)\) \Comment Substitution step (\texttt{xAXPY})

    \EndFor

    \EndIf

    \EndProcedure

  \end{algorithmic}
\end{algorithm}
\noindent
In the best-case scenario, i.e.\ if \(U\) is well-conditioned, forming
the initial growth bounds is the only additional work compared to
standard back substitution. Note that these bounds require computing
norms of the columns of \(U\), excluding the diagonal. In the
worst-case scenario, rescaling \(b\) at each step in back substitution
will triple the flop count. However, we find in practice that
rescaling is a relatively rare event, even in ill-conditioned and
singular matrices.

\paragraph{} Safeguarded back substitution is easily generalized to
the safe multi-shift triangular solve problem. Given nonzero
right-hand side vectors \(b_1,\cdots,b_n\) and shifts
\(\lambda_1,\cdots,\lambda_n\), we seek nonzero solution vectors
\(x_1,\cdots,x_n\) and scaling factors \(s_1,\cdots,s_n\) in
\(\left[0,1\right]\) such that
\begin{align}
  \left( U - \lambda_j I\right) x_j &= s_j b_j.
\end{align}
For safeguarded, blocked back substitution, we just need to ensure
that the matrix multiplications in the substitution step do not cause
numerical issues:
\begin{algorithm}[H]
  \caption{Safe multi-shift triangular solve with safeguarded, blocked back substitution }
  \label{algorithm:safe multi-shift trsm}
  \begin{algorithmic}
    \Procedure{SafeMultiShiftTrsm}{$U, \lambda, B, s $}
    \Comment \(B\) is overwritten with \(X\)

    \State \(s := \left[1, \cdots, 1 \right]^T\) \Comment \(n\) entries

    \For{$j=1:n$}

    \State \( G(j) := \left\lVert B(:,j) \right\rVert_\infty\)

    \EndFor

    \For{$i = m-n_b+1:-n_b:1$} \Comment Assume \(m\) is a multiple of
    \(n_b\)

    \State \( \mathcal{I}_0 := 1:i-1 \)

    \State \( \mathcal{I}_1 := i:i+n_b-1\)

    \State \( \mathcal{I}_2 := i+n_b:m\)
    
    \For{$j=1:n$}

    \State \Call{SafeTrsv}{$U(\mathcal{I}_1,\mathcal{I}_1)-\lambda(j) * I, B(\mathcal{I}_1,j), t$} 
    \Comment Diagonal block step

    \If{$t<1$}

    \State \(B(\mathcal{I}_0,j) := t*B(\mathcal{I}_0,j)\) \Comment \texttt{xSCAL}

    \State \(B(\mathcal{I}_2,j) := t*B(\mathcal{I}_2,j)\) \Comment \texttt{xSCAL}

    \State \(s(j) := t * s(j)\)

    \State \(G(j) := t * G(j)\)

    \EndIf

    \EndFor

    \For{$j=1:n$}

    \State \(G(j) := G(j) + \sum_{k\in \mathcal{I}_1} \left\lVert U(\mathcal{I}_0,k) \right\rVert_{\infty} \left\lVert B(\mathcal{I}_1,j) \right\rVert_\infty \)

    \If{$G(j) \geq \Omega$}

    \State Choose \(t\in\left(0,1\right)\) so that \(t * G(j) < \Omega\)

    \State \(B(:,j) := t * B(:,j)\) \Comment \texttt{xSCAL}

    \State \(s(j) := t* s(j)\)

    \State \(G(j) := t * G(j)\)    

    \EndIf

    \EndFor

    \State \( B(\mathcal{I}_0,:) := B(\mathcal{I}_0,:) - U(\mathcal{I}_0,\mathcal{I}_1) * B(\mathcal{I}_1,:) \)
    \Comment Substitution step (\texttt{xGEMM})

    \EndFor

    \EndProcedure
  \end{algorithmic}
\end{algorithm}
\noindent
Note that each application of \textsc{SafeTrsv} requires norms of the
columns of \(U(\mathcal{I}_1,\mathcal{I}_1)\), excluding the diagonal,
in order to form growth bounds. Thus, we can improve performance in
the diagonal block step by reusing this data for each right-hand side.

\section{Eigenvector Computation} \label{section:eigenvector}
We shall now apply a safe multi-shift triangular solve to compute the
eigenvectors of a general \(n\times n\) matrix \(A\).  Assuming that
\(A\) is nondefective, we seek an eigenvalue decomposition
\(A=X\Lambda X^{-1}\) where \(\Lambda\) is a diagonal eigenvalue
matrix and \(X\) an eigenvector matrix.\footnote{ If \(A\) is
  defective, a nondefective matrix can be obtained with a small
  perturbation of \(A\).}  We begin by computing the Schur
decomposition \(A=Q TQ^H\) where \(T\) is upper triangular and \(Q\)
unitary. In LAPACK's general eigensolver routine \texttt{xGEEV}, this
is performed efficiently with Level 3 BLAS by converting \(A\) to
upper Hessenberg form (\texttt{xGEHRD}) \cite{quintana2006improving},
converting Householder transforms to a unitary matrix
(\texttt{xUNGHR}), and applying the QR algorithm (\texttt{xHSEQR})
\cite{braman2002multishift1, braman2002multishift2}. Since similar
matrices have identical eigenvalues, \(\Lambda\) can be obtained by
simply reading off the diagonal of \(T\).  Now, all that remains is to
find a triangular eigenvector matrix \(Z\) such that \(T=Z \Lambda
Z^{-1}\) and to compute the back substitution \(X=QZ\).  Assuming that
\(Z\) is upper triangular and letting \(\lambda_k\) and \(z_k\)
respectively denote the \(k\)th eigenvalue and triangular eigenvector,
we require
\begin{align}
  \left[
  \begin{matrix}
    T_{11} & u & T_{13} \\
    0 & \lambda_k & v^T \\
    0 & 0 & T_{33}
  \end{matrix}
  \right]
  \left[
  \begin{matrix}
    \hat{z} \\ s \\ 0
  \end{matrix}
  \right]
  &=
  \lambda_k
  \left[
  \begin{matrix}
    \hat{z} \\ s \\ 0
  \end{matrix}
  \right].
\end{align}
This system is satisfied if and only if \(\hat{z}\) is a solution to a
\(k-1\times k-1\) shifted triangular system,
\begin{align}
  \left(T_{11} - \lambda_k I\right) \hat{z} &= -s u.
  \label{eq:triangeig-2}
\end{align}
This is similar to the form of the safe multi-shift triangular solve
problem, but each triangular system has a different size. LAPACK's
triangular eigenvector routine \texttt{xTREVC} approaches this problem
by calling \texttt{xLATRS} for each triangular eigenvector and back
transforming with calls to \texttt{xGEMV}. These routines are limited
to Level 2 BLAS and hence achieve poor performance. On multicore
architectures, this procedure can be accelerated by performing
\texttt{xLATRS } in parallel and by blocking the back substitution
into Level 3 BLAS \texttt{xGEMM} calls
\cite{gates2014accelerating}. However, a hardware-independent solution
is preferable for the sake of portability.  Observe that
(\ref{eq:triangeig-2}) holds if and only if
\begin{align}
  \left(\left[
    \begin{matrix}
      T_{11} & u & T_{13} \\
      0 & \lambda_k & v^T \\
      0 & 0 & T_{33}
    \end{matrix}
  \right] - \lambda_k I\right)
  \left[
    \begin{matrix}
      \hat{z} \\ 0 \\ 0
    \end{matrix}
  \right]
  &=
  s \left[
    \begin{matrix}
      -u \\ 0 \\ 0
    \end{matrix}
  \right].
\end{align}
This is the form for a safe multi-shift triangular solve.\footnote{If
  we apply back substitution to a right-hand vector where the last
  \(n-k+1\) entries are zero, the corresponding entries in the
  solution will also be zero.} Furthermore, the right-hand side matrix
will be strictly upper triangular, so we can shortcut the back
substitution to avoid unnecessary flops involving zeros. After
computing a safe multi-shift triangular solve, the triangular
eigenvectors are obtained by putting the scaling factors on the
diagonal of the solution matrix. The algorithm is outlined below:
\begin{algorithm}[H]
  \caption{Triangular and general eigensolvers}
  \label{algorithm:eig}
  \begin{algorithmic}
    
    \Function{TriangEig}{$T$}

    \State \(\lambda := \text{diag}(T)\)

    \State \(Z := -T\)

    \State \(\text{diag}(Z) := 0\)

    \State \(s := \left[1,\cdots,1\right]^T\)

    \State \Call{SafeMultiShiftTrsm}{$T, \lambda, Z, s$}
    \Comment Shortcutted to exploit structure

    \State \(\text{diag}(Z) := s\)

    \State \Return \(\lambda, Z\)

    \EndFunction

    \Function{Eig}{$A$}

    \State Compute Schur decomposition \(A=QTQ^H\) \Comment \texttt{xGEHRD}, \texttt{xUNGHR}, \texttt{xHSEQR}

    \State \(\lambda, Z := \) \Call{TriangEig}{$T$}

    \State \(X := Q Z\) \Comment \texttt{xGEMM}

    \State \Return \(\lambda, X\)

    \EndFunction

  \end{algorithmic}
\end{algorithm}
\noindent
Since the QR algorithm and safe multi-shift triangular solve are both
backward stable, this method is also backward stable. Thus, we can
utilize bounds on the condition number of the eigenvector problem
(LAPACK routines \texttt{xTRSNA} and \texttt{xTRSEN})
\cite{bai1989conditioning} to compute error bounds for the
eigenvectors.

\section{Pseudospectra Computation}
The classical approach to study the behavior of an \(n\times n\)
matrix \(A\) is to analyze the distribution of its eigenvalues. The
eigenvalues can can be defined in terms of the matrix resolvent
\(f_A(z)=\left(z I- A\right)^{-1}\),
\begin{align}
  \Lambda(A) &= \left\lbrace z\in\mathbb{C} : \text{\(\left( zI
        -A\right)^{-1}\) is unbounded} \right\rbrace.
\end{align}
However, eigenvalues may be insufficient to explain the behavior of
\(A\)
if it is highly nonnormal.  In particular, small perturbations to
\(A\)
can dramatically change the eigenvalue distribution. To capture this
nonnormal behavior, it is preferable to analyze the pseudospectra of
\(A\).
Given \(\epsilon>0\),
the \(p\)-norm \(\epsilon\)-pseudospectrum of \(A\) is defined as
\begin{align}
  \Lambda_\epsilon^p(A) &= \left\lbrace z\in\mathbb{C} : \left\lVert
      \left(zI-A\right)^{-1}\right\rVert_p \geq \frac{1}{\epsilon}
  \right\rbrace.
\end{align}
A full theory of pseudospectra is developed in
\cite{trefethen2005spectra}. We focus on the case \(p=2\),
although similar results can be obtained with \(p=1\)
using a more sophisticated algorithm
\cite{higham2000block}.\footnote{The Hager/Higham algorithm for
  computing 1-norm pseudospectra is similar to the Van Loan/Lui
  algorithm discussed below in that it involves establishing a grid in
  the complex plane, computing a Schur decomposition, and performing
  shifted triangular solves with grid points as shifts. Thus, it can
  be similarly accelerated with blocked multi-shift triangular
  solves. }

\paragraph{} Pseudospectra are typically computed by establishing a
grid with \(N\) points on a region of the complex plane, computing the
resolvent norm \(\lVert (zI - A)^{-1} \rVert_2\) at each grid point
\(z\), and visualizing with a contour plotter. Letting
\(\sigma_\text{max}(\cdot)\) and \(\sigma_\text{min}(\cdot)\) denote
the largest and smallest singular values of an input matrix,
respectively, we remark that the resolvent norm satisfies
\begin{align}
  \left\lVert \left(zI - A\right)^{-1} \right\rVert_2
  &= \sigma_\text{max}\left( \left(zI - A\right)^{-1} \right) \nonumber \\
  &= \frac{1}{\sigma_\text{min}\left( zI - A \right)}.
\end{align}
Thus, one could naively compute pseudospectra by computing the SVD of
\(zI-A\)
for each grid point \(z\)
and reporting the reciprocal of the smallest singular value. However,
this involves a total of \(O(Nn^3)\)
flops, which is prohibitively expensive for large matrices unless the
grid is very coarse. The Van Loan/Lui algorithm improves on the
computational cost by proceeding in two stages \cite{van1984near,
  lui1997computation}.\footnote{Most authors attribute this algorithm
  to Lui, but the algorithm was presented (in a different context) by
  Van Loan more than a decade earlier.} It begins by computing a Schur
decomposition \(A=QTQ^H\)
where \(T\)
is upper triangular and \(Q\)
unitary (see discussion in Section \ref{section:eigenvector}). Since
matrix norms are invariant under unitary transformations,
\begin{align*}
  \left\lVert \left( zI-A \right)^{-1} \right\rVert_2
  &= \left\lVert \left( zQQ^H - QTQ^H \right)^{-1} \right\rVert_2 \\
  &= \left\lVert -Q\left( T -zI\right)^{-1} Q^H \right\rVert_2 \\
  &= \left\lVert \left( T-zI \right)^{-1}\right\rVert_2 \\
  &= \sigma_\text{max}\left( \left( T-zI \right)^{-1} \right).
\end{align*}
Letting \(\lambda_\text{max}(\cdot)\) denote the largest eigenvalue of
a Hermitian matrix,
\begin{align}
  \left\lVert \left( zI-A \right)^{-1} \right\rVert_2
  &= \sqrt{\lambda_\text{max}\left( \left( T-zI \right)^{-H} \left( T-zI \right)^{-1} \right)}.
\end{align}
A Krylov eigensolver like the Lanczos algorithm can estimate the
largest eigenvalue of \(\left( T-zI \right)^{-H} \left( T-zI
\right)^{-1}\) by repeatedly applying it to a vector. Each matrix
product can be computed with two triangular solves, for a total cost
of \(O(n^2)\) flops. The algorithm is outlined below:
\begin{algorithm}[H]
  \caption{Van Loan/Lui algorithm}
  \label{algorithm:van loan/lui}
  \begin{algorithmic}
    \Function{VanLoanLui}{$A, z$}

    \State Compute Schur decomposition \(A=Q T Q^H\) \Comment \texttt{xGEHRD}, \texttt{xUNGHR}, \texttt{xHSEQR}
    
    \For{$i=1:N$}

    \State \(B := \left(T-z(i)*I\right)^{-H} \left(T-z(i)*I\right)^{-1}\)
    \Comment Not formed explicitly

    \State \(\lambda(i) := \lambda_\text{max}\left( B \right)\)
    \Comment Krylov eigensolver

    \State \(r(i) := \sqrt{\lambda(i)}\)

    \EndFor

    \State \Return \(r\) \Comment Visualize with contour plotter

    \EndFunction
  \end{algorithmic}
\end{algorithm}
\noindent
This algorithm takes \(O(n^3+Nn^2)\) flops and has been implemented in
the popular Matlab package EigTool \cite{wright2002eigtool}. However,
we make the additional observation that treating each grid point as a
shift puts the triangular solves in the form of a multi-shift
triangular solve. We can thus achieve Level 3 BLAS performance in both
the first and second stage of the Van Loan/Lui algorithm.

\section{Implementation}
Although the algorithms discussed so far are sequential, they are
readily parallelizable on distributed-memory architectures. For
instance, if matrix data for a multi-shift triangular solve is stored
element-wise across multiple processes, the diagonal block step can be
performed (redundantly) on each process and the substitution step can
be performed with a distributed matrix product.  Sequential and
parallel versions of the above algorithms are implemented as part of
Elemental, an open-source C++ library for distributed-memory linear
algebra and optimization \cite{poulson2013elemental}. Several relevant
functions in Elemental are listed below:
\begin{itemize}
\item \texttt{MultiShiftTrsm} - Multi-shift triangular solve.
\item \texttt{SafeMultiShiftTrsm} - Safe multi-shift triangular solve.
\item \texttt{TriangEig} - Triangular eigensolver.
\item \texttt{Eig} - General eigensolver.
\item \texttt{SpectralCloud} - Resolvent norm with user-specified complex shifts.
\item \texttt{SpectralWindow} - Resolvent norm over user-specified grid in complex plane.
\item \texttt{SpectralPortrait} - Resolvent norm over automatically-determined grid in complex plane.
\end{itemize}

\section{Experimental Results}
\subsection{Multi-shift Triangular Solve}
\begin{figure}[H]
  \centering
  \epsfig{file=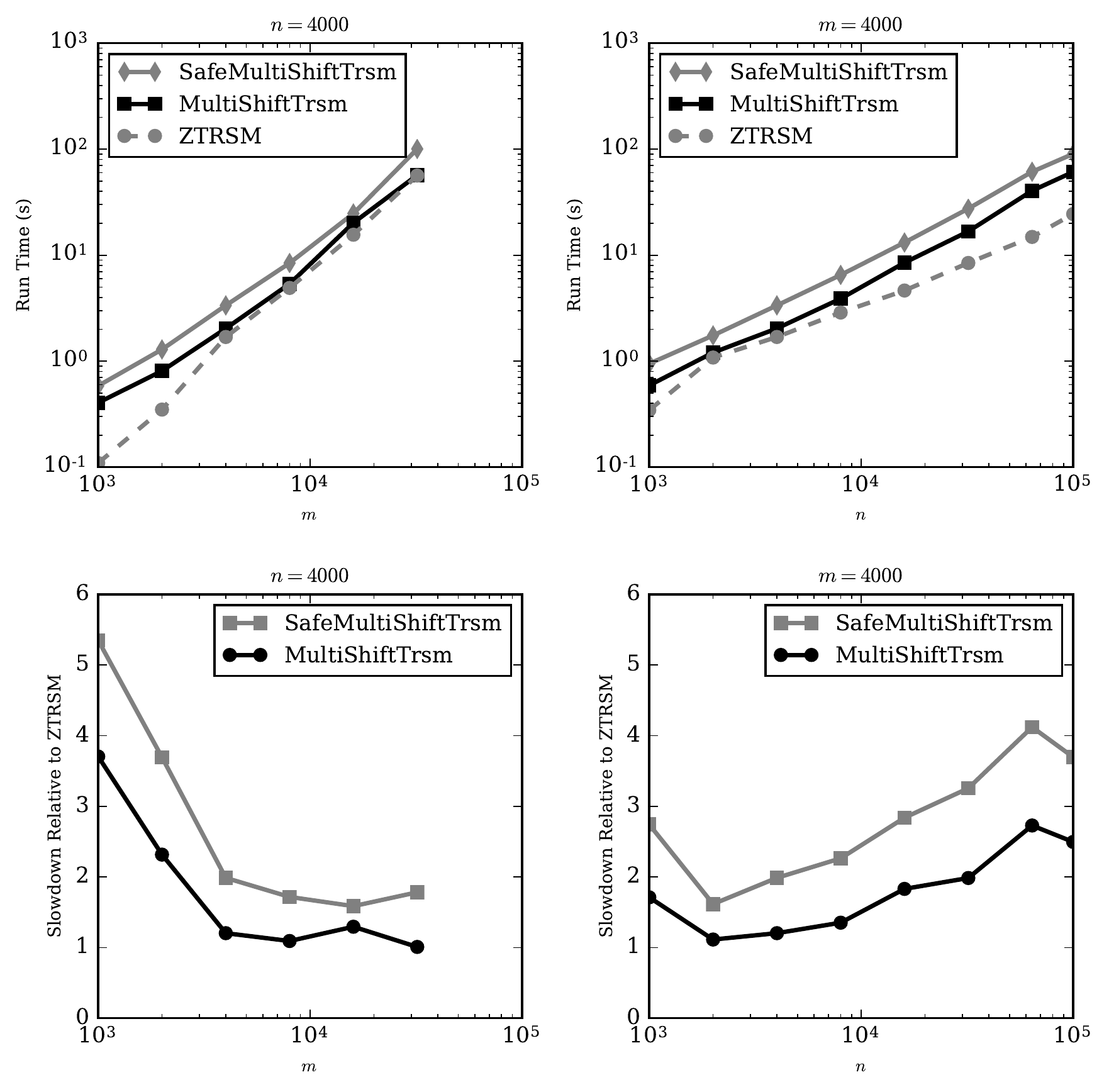, width=0.8\linewidth}
  \caption{Scaling study of triangular solve, multi-shift triangular
    solve, and safe multi-shift triangular solve. Matrices were
    generated by finding Hermitian matrices with uniform random
    eigenvalues in the interval \(\left[1,2\right]\) and deleting
    entries in the strict lower triangle. Shifts were drawn uniform
    randomly from the ball \(B\left(1.5,0.5\right)\).}
  \label{fig:mstrsm}
\end{figure}

Numerical experiments with the multi-shift triangular solve and safe
multi-shift triangular solve were performed with two 4-core Intel
Haswell Xeon E5-2623 v3 CPUs at 3.00 GHz. All calculations were
performed with double-precision complex numbers and BLAS calls were
performed with Intel MKL. A scaling study of the multi-shift
triangular solve and safe multi-shift triangular is presented in
Figure \ref{fig:mstrsm}. The safe multi-shift triangular solve was
consistently slower than the standard multi-shift triangular solve,
which was in turn slower than the Level 3 BLAS triangular solve
routine \texttt{ZTRSM}. The performance of the multi-shift triangular
solve and safe multi-shift triangular solve were especially poor when
the number of right-hand sides was much larger than the matrix
dimension, i.e.\ \(m\ll n\). However, when \(m\geq n\), we see that
the performance of the multi-shift triangular solve is typically
within a factor of 1.5 of Level 3 BLAS performance and that the safe
multi-shift triangular solve is typically within a factor of 2.

\subsection{Eigenvector Computation}
\begin{figure}[H]
  \centering
  \begin{tabular}{c}
    \epsfig{file=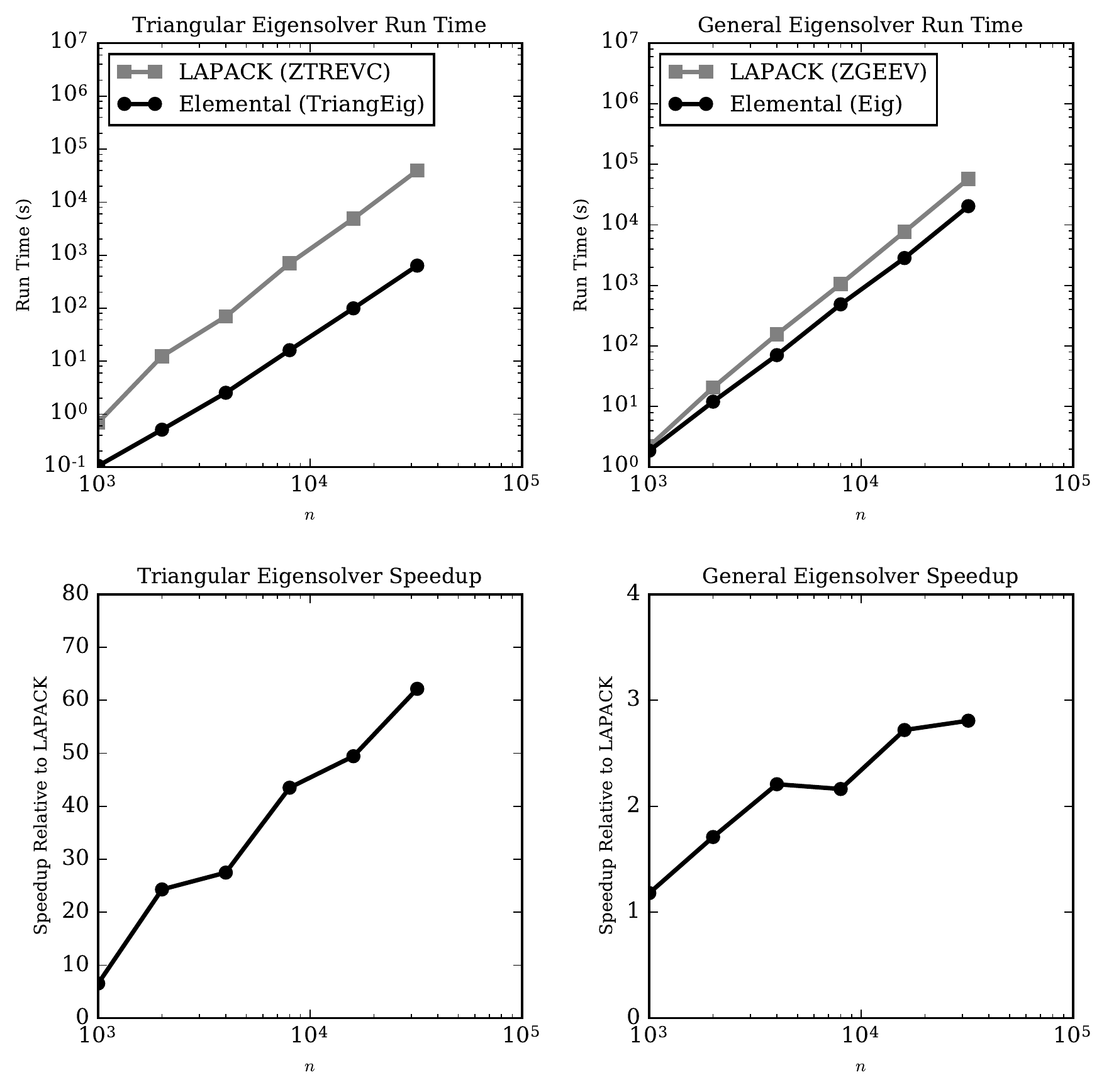, width=0.8\linewidth}
  \end{tabular}
  \caption{Timing experiments for triangular and general
    eigensolvers. Matrices were generated by choosing entries uniform
    randomly from the unit ball.}
  \label{fig:unif}
\end{figure}

\begin{figure}[H]
  \centering
  \epsfig{file=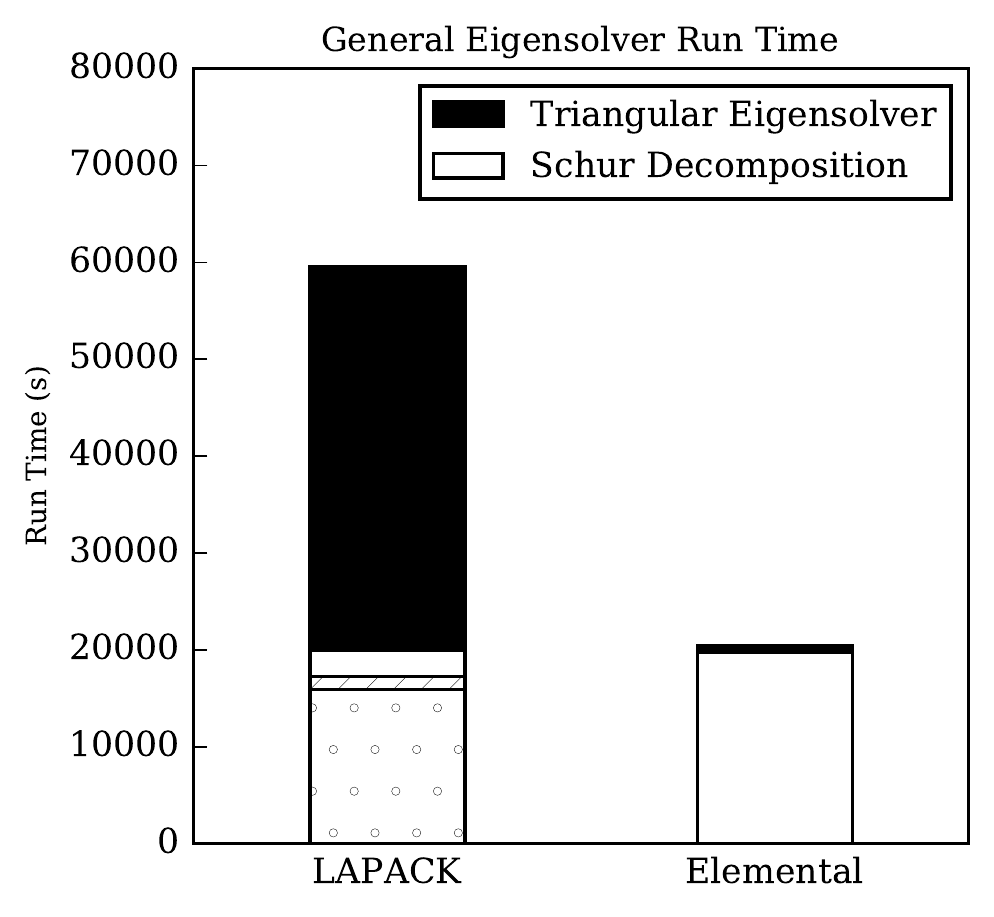, width=0.4\linewidth}
  \caption{Timings for the general eigensolvers in LAPACK
    (\texttt{ZGEEV}) and Elemental (\texttt{Eig}).  A \(32000\times
    32000\) matrix was generated by choosing entries uniform randomly
    from the unit ball. The timing for LAPACK's Schur decomposition is
    subdivided into three stages: \texttt{ZGEHRD} (dots),
    \texttt{ZUNGHR} (stripes), \texttt{ZHSEQR} (no pattern). The
    timing for Elemental's triangular eigensolver includes back
    transformation.  }
  \label{fig:unif-parts}
\end{figure}

\begin{figure}[H]
  \centering
  \begin{tabular}{c}
    \epsfig{file=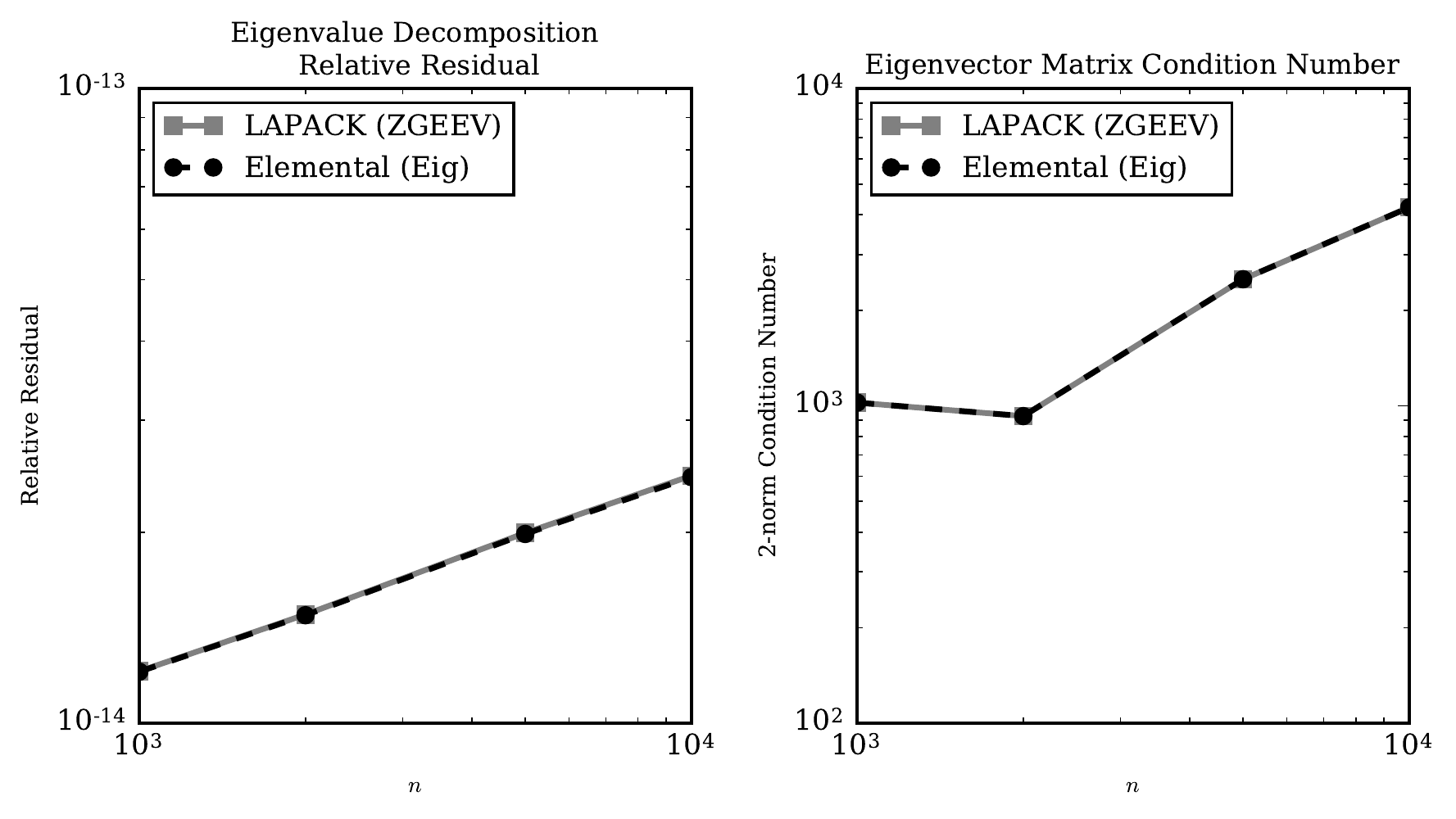, width=0.8\linewidth}
  \end{tabular}
  \caption{Scaling study for triangular and general
    eigensolvers. Matrices were generated by choosing entries uniform
    randomly from the unit ball.}
  \label{fig:unif-err}
\end{figure}

Timing experiments with the triangular and general eigensolvers were
performed with the same setup as above (two 4-core Intel Haswell Xeon
E5-2623 v3 CPUs at 3.00 GHz, double-precision complex data type, Intel
MKL).  Scaling studies of the triangular eigensolver and general
eigensolver are shown in Figure \ref{fig:unif}. The triangular
eigensolver appears to be asymptotically faster than the LAPACK
routine \texttt{ZTREVC} and we report a speedup by a factor of 60 on a
\(32000\times 32000\) matrix. The general eigensolver appears to
converge to a threefold speedup relative to the LAPACK routine
\texttt{ZGEEV}. If we inspect the time spent in each stage of the
calculation, as shown in Figure \ref{fig:unif-parts}, we see that
Elemental and LAPACK take nearly the same amount of time to compute a
Schur decomposition. However, roughly two thirds of LAPACK's run time
takes place in the triangular eigensolver while Elemental's triangular
eigensolver takes a negligible amount of time.
The eigenvalue decomposition \(A=X\Lambda X^{-1}\) can be validated by
computing the relative residual \({{\left\lVert A X - X \Lambda
    \right\rVert}_F}/{ {\left\lVert A \right\rVert}_F }\) and the
2-norm condition number \(\left\lVert X\right\rVert_2 \left\lVert
  X^{-1}\right\rVert_2\). As shown in Figure
\ref{fig:unif-err},\footnote{These experiments were performed with a
  4-core Intel Nehalem Core i7-870 CPU at 2.93 GHz, also using
  double-precision complex data type and Intel MKL.}  Elemental yields
nearly identical results to LAPACK. In particular, the relative
residuals are smaller than \(10^{-13}\) and the condition numbers are
not singular, suggesting that both eigensolvers achieve reasonable
quality.

\subsection{Pseudospectra Computation}
\begin{figure}[H]
  \centering
  \begin{tabular}{ccc}
    \epsfig{file=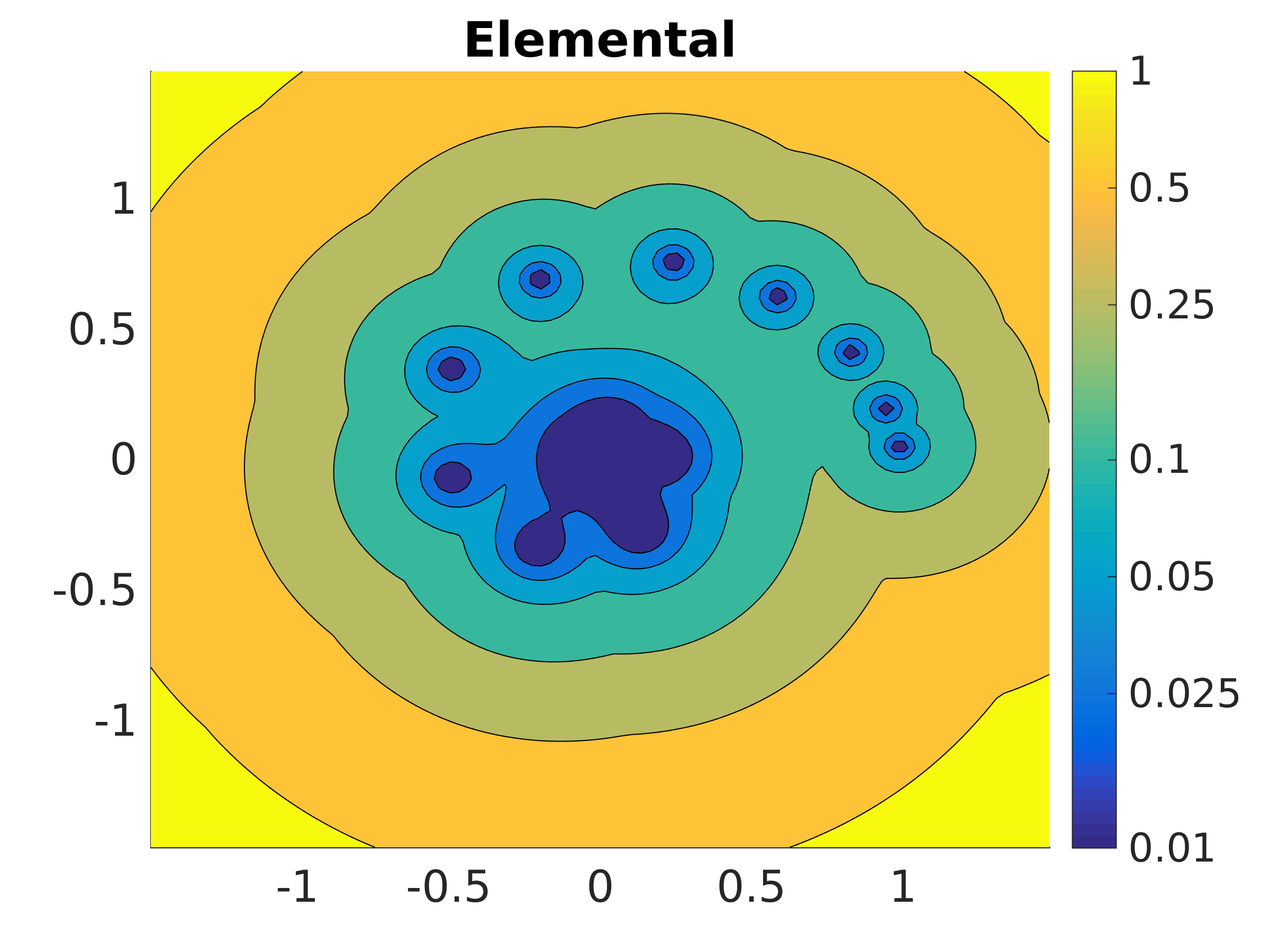, width=0.3\linewidth}
    & \epsfig{file=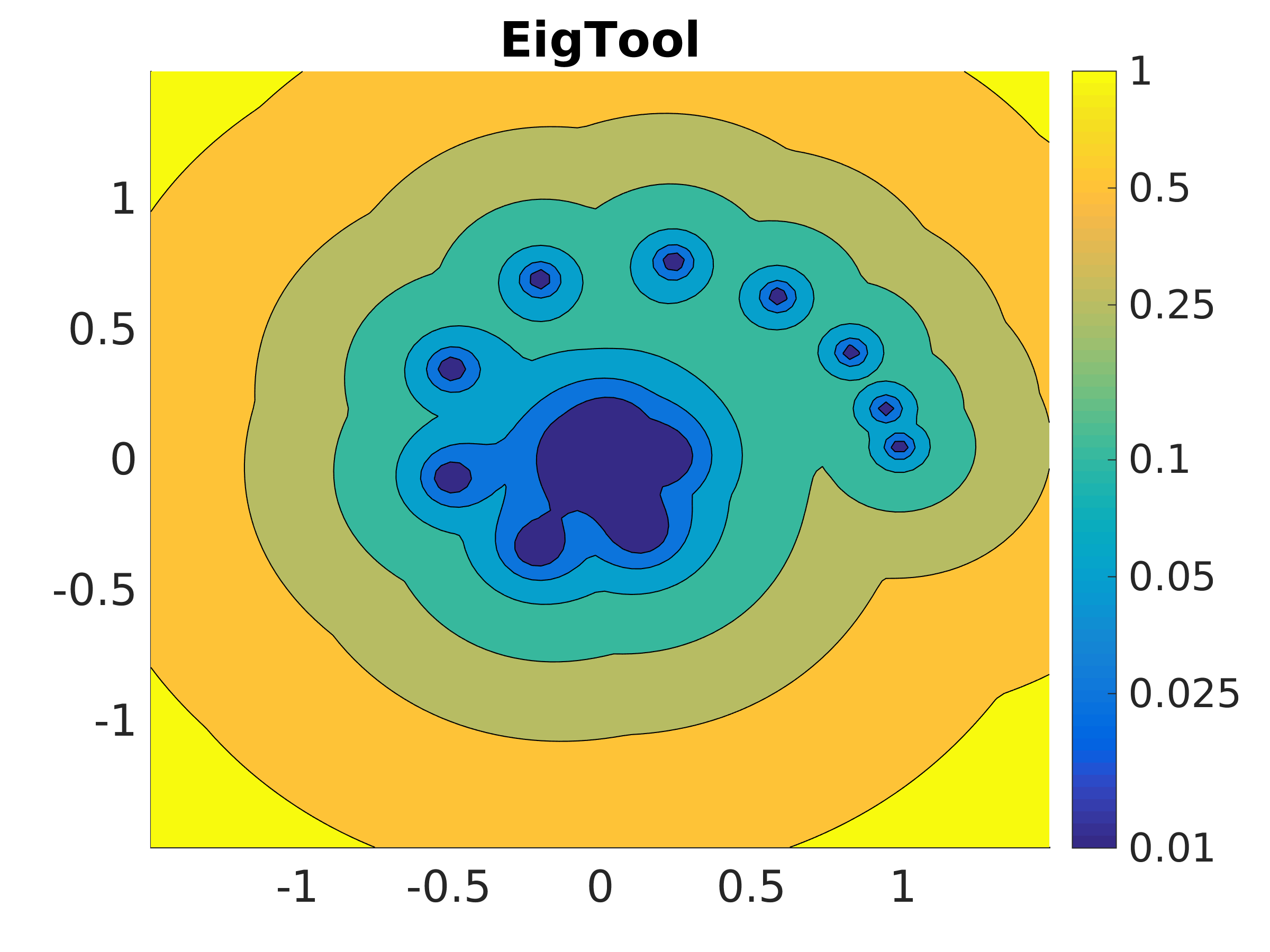, width=0.3\linewidth}
    & \epsfig{file=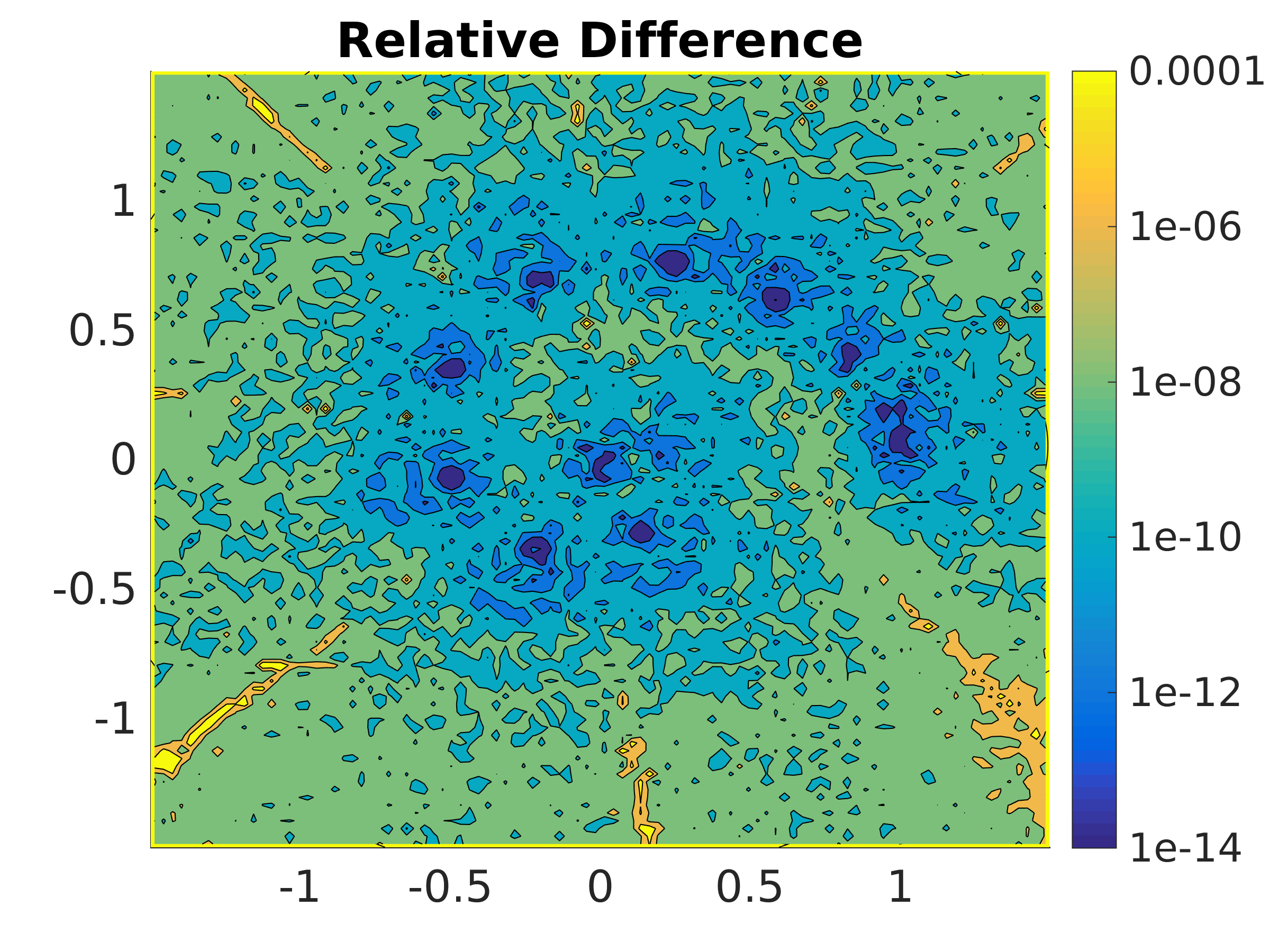, width=0.3\linewidth} \\
    \textbf{(a)} & \textbf{(b)} & \textbf{(c)}
  \end{tabular}
  \caption{\(\epsilon\)-pseudospectra with several values of
    \(\epsilon\) and the relative difference in computed resolvent
    norm.  The matrix is a \(100\times 100\) discretization of the
    Fox/Li operator with parameter \(F=10\) and the image is
    constructed from the resolvent norm computed at 10000 grid points
    \cite{fox1961resonant}. See Chapter 60 of
    \cite{trefethen2005spectra} for a discussion on the pseudospectra
    of the Fox/Li operator.}
  \label{fig:ps_plots}
\end{figure}

\begin{figure}[H]
  \centering
  \epsfig{file=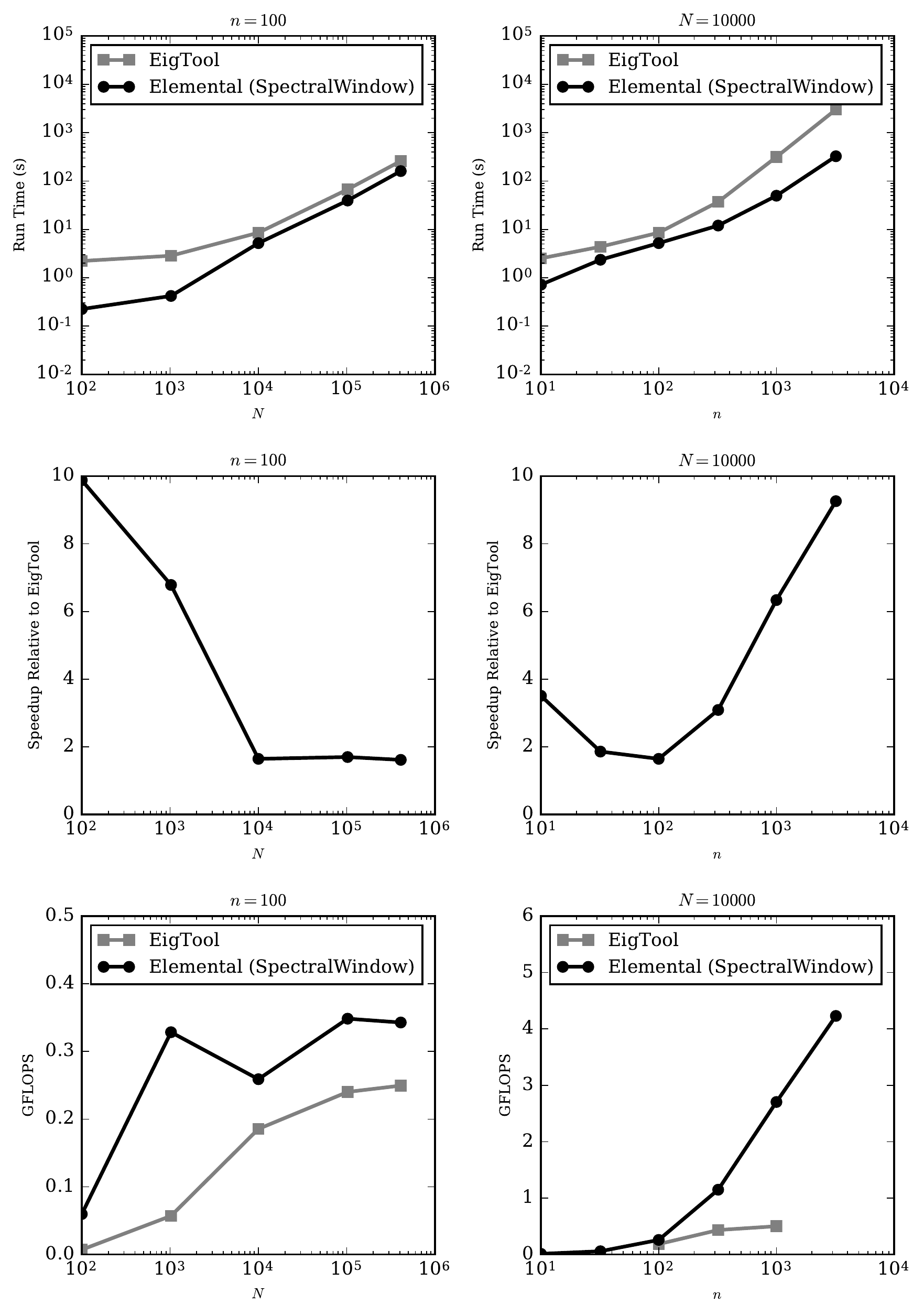, width=0.8\linewidth}
  \caption{Scaling study for pseudospectra computation on
    discretizations of the Fox/Li operator with parameter \(F=10\)
    \cite{fox1961resonant}.}
  \label{fig:ps_timing}
\end{figure}

Experiments with pseudospectra solvers were performed with a 4-core
Intel Nehalem Core i7-870 CPU at 2.93 GHz. Calculations were performed
with double-precision complex numbers and BLAS calls were performed
with Intel MKL (for both Elemental and MATLAB). Pseudospectra computed
by Elemental and EigTool are qualitatively identical, as shown in
Figure \ref{fig:ps_plots}.
Scaling studies of pseudospectra computation are shown in Figure
\ref{fig:ps_timing}. Given sufficiently many grid points, we see that
Elemental achieves a speedup by a factor of 1.6 when computing
pseudospectra of a \(100\times 100\) matrix. This modest result can be
explained by noting that Krylov eigensolvers like the Lanczos method
or Arnoldi method require tridiagonal or Hessenberg
eigensolvers.\footnote{EigTool uses the Lanczos method for
  sufficiently large matrices and Elemental implements a variety of
  eigensolvers. Our experiments with Elemental use the implicitly
  restarted Arnoldi method.} When the matrix dimension is small, these
eigensolvers take a non-trivial portion of the computation. However,
when the matrix dimension is large, the computation is dominated by
triangular solves. In this regime, Elemental's pseudospectra solver
fully exploits Level 3 BLAS performance and achieves a ninefold
speedup when computing the resolvent norm of a \(3200\times 3200\)
matrix at \(10000\) grid points.

\section{Conclusion}
We have shown that multi-shift triangular solves can be performed
efficiently by exploiting Level 3 BLAS routines. Numerical experiments
suggest that using a safe multi-shift triangular solve to compute
triangular eigenvectors is substantially faster than the algorithm
presently used in LAPACK, yielding a 60x speedup when computing the
eigenvectors of a triangular matrix and a 3x speedup for a general
matrix. Pseudospectra computation with multi-shift triangular solves
also appears to exhibit a ninefold speedup relative to EigTool when
the matrix dimension is sufficiently large.  Future work will include
real arithmetic multi-shift triangular solves and Fortran
implementations for inclusion into LAPACK.

\section{Acknowledgments}
Tim Moon was supported by a Simons Graduate Research Assistantship and
Jack Poulson by AFRL Contract FA8750-12-2-0306.

\appendix
\section{Generalized Multi-Shift Triangular Solve}
The multi-shift triangular solve problem can be generalized and
applied to compute generalized eigenvectors. Given \(m\times m\) upper
triangular matrices \(U\) and \(V\), right-hand side vectors
\(b_1,\cdots,b_n\), and shifts \(\lambda_1,\cdots,\lambda_n\), we seek
solution vectors \(x_1,\cdots,x_n\) such that
\begin{align}
  \left( U - \lambda_j V \right) x_j &= b_j .
\end{align}
This problem can be solved with small modifications to Algorithm
\ref{algorithm:multi-shift trsm}:
\begin{algorithm}[H]
  \caption{Generalized multi-shift triangular solve with blocked back substitution}
  \label{algorithm:generalized multi-shift trsm}
  \begin{algorithmic}
    \Procedure{GeneralizedMultiShiftTrsm}{$U, V, \lambda, B $}
    \Comment \parbox[t]{0.22\linewidth}{\(B\) is overwritten with \(X\)}

    \For{$i = m-n_b+1:-n_b:1$} \Comment Assume \(m\) is a multiple of
    \(n_b\)

    \State \( \mathcal{I}_0 := 1:i-1 \)

    \State \( \mathcal{I}_1 := i:i+n_b-1\)
    
    \For{$j=1:n$}

    \State \(U' := U(\mathcal{I}_1,\mathcal{I}_1)-\lambda(j) * V(\mathcal{I}_1,\mathcal{I}_1)\) \Comment \texttt{xAXPY}

    \State \Call{Trsv}{$U', B(\mathcal{I}_1,j)$} 
    \Comment \texttt{xTRSV}

    \State \( C(:,j) := \lambda(j) * B(\mathcal{I}_1,j)\) \Comment \texttt{xSCAL}

    \EndFor

    \State \( B(\mathcal{I}_0,:) := B(\mathcal{I}_0,:) - U(\mathcal{I}_0,\mathcal{I}_1) * B(\mathcal{I}_1,:) \)
    \Comment \texttt{xGEMM}

    \State \( B(\mathcal{I}_0,:) := B(\mathcal{I}_0,:) + V(\mathcal{I}_0,\mathcal{I}_1)* C \) \Comment \texttt{xGEMM}

    \EndFor

    \EndProcedure
  \end{algorithmic}
\end{algorithm}
\noindent
Making similar changes to Algorithm \ref{algorithm:safe multi-shift
  trsm} yields a robust algorithm:
\begin{algorithm}[H]
  \caption{Safe, generalized multi-shift triangular solve with
    safeguarded, blocked back substitution }
  \label{algorithm:safe generalized multi-shift trsm}
  \begin{algorithmic}
    \Procedure{SafeGeneralizedMultiShiftTrsm}{$U, V, \lambda, B, s $}
    \Comment \parbox[t]{0.13\linewidth}{\(B\) is overwritten with \(X\)}

    \State \(s := \left[1, \cdots, 1 \right]^T\) \Comment \(n\) entries

    \For{$j=1:n$}

    \State \( G(j) := \left\lVert B(:,j) \right\rVert_\infty\)

    \EndFor

    \For{$i = m-n_b+1:-n_b:1$} \Comment Assume \(m\) is a multiple of
    \(n_b\)

    \State \( \mathcal{I}_0 := 1:i-1 \)

    \State \( \mathcal{I}_1 := i:i+n_b-1\)

    \State \( \mathcal{I}_2 := i+n_b:m\)
    
    \For{$j=1:n$}

    \State \(U' := U(\mathcal{I}_1, \mathcal{I}_1) - \lambda(j) * V(\mathcal{I}_1, \mathcal{I}_1)\) \Comment\texttt{xAXPY}

    \State \Call{SafeTrsv}{$U', B(\mathcal{I}_1,j), t$} 
    \Comment Diagonal block step

    \If{$t<1$}

    \State \(B(\mathcal{I}_0,j) := t*B(\mathcal{I}_0,j)\) \Comment \texttt{xSCAL}

    \State \(B(\mathcal{I}_2,j) := t*B(\mathcal{I}_2,j)\) \Comment \texttt{xSCAL}

    \State \(s(j) := t * s(j)\)

    \State \(G(j) := t * G(j)\)

    \EndIf

    \State \(C(:,j) := \lambda(j) * B(\mathcal{I}_1,j)\) \Comment \texttt{xSCAL}

    \EndFor

    \For{$j=1:n$}

    \State \(G(j) := G(j) + \sum_{k\in \mathcal{I}_1} \left( \left\lVert U(\mathcal{I}_0,k) \right\rVert_{\infty} + \left| \lambda(j)\right|  {\left\lVert V(\mathcal{I}_0,k)\right\rVert}_\infty \right) \left\lVert B(\mathcal{I}_1,j) \right\rVert_\infty \)

    \If{$G(j) \geq \Omega$}

    \State Choose \(t\in\left(0,1\right)\) so that \(t * G(j) < \Omega\)

    \State \(B(:,j) := t * B(:,j)\) \Comment \texttt{xSCAL}

    \State \(s(j) := t* s(j)\)

    \State \(G(j) := t * G(j)\)    

    \EndIf

    \EndFor

    \State \( B(\mathcal{I}_0,:) := B(\mathcal{I}_0,:) - U(\mathcal{I}_0,\mathcal{I}_1) * B(\mathcal{I}_1,:) \)
    \Comment \texttt{xGEMM}

    \State \( B(\mathcal{I}_0,:) := B(\mathcal{I}_0,:) + V(\mathcal{I}_0,\mathcal{I}_1) * C \)
    \Comment \texttt{xGEMM}

    \EndFor

    \EndProcedure
  \end{algorithmic}
\end{algorithm}
\noindent
This routine can be used to solve the generalized eigenvalue problem:
\begin{algorithm}[H]
  \caption{Generalized, triangular and generalized, general
    eigensolvers}
  \label{algorithm:eig}
  \begin{algorithmic}
    
    \Function{GeneralizedTriangEig}{$T,S$}

    \State \(\lambda := \text{diag}(T) ./ \text{diag}(S)\)

    \State \(Z := -T + S * \text{diag}(\lambda)\)

    \State \(\text{diag}(Z) := 0\)

    \State \(s := \left[1,\cdots,1\right]^T\)

    \State \Call{SafeGeneralizedMultiShiftTrsm}{$T,S, \lambda, Z, s$}

    \State \(\text{diag}(Z) := s\)

    \State \Return \(\lambda, Z\)

    \EndFunction

    \Function{GeneralizedEig}{$A$}

    \State Compute generalized Schur decomposition \(A=QTP^H, B=QSP^H\)

    \State \(\lambda, Z := \) \Call{GeneralizedTriangEig}{$T,S$}

    \State \(X := P Z\) \Comment \texttt{xGEMM}

    \State \Return \(\lambda, X\)

    \EndFunction

  \end{algorithmic}
\end{algorithm}

\bibliography{main}{}
\bibliographystyle{plain}

\end{document}